\documentclass[conference]{IEEEtran}
\usepackage{cite}
\usepackage{amsmath,amssymb,amsfonts}
\usepackage{algorithmic}
\usepackage{graphicx}
\usepackage{textcomp}
\usepackage{xcolor}
\usepackage{algorithm}
\usepackage{hyperref}
\usepackage[version=4]{mhchem}

\newcommand{\bra}[1]{\left\langle #1 \right|}
\newcommand{\ket}[1]{\left| #1 \right\rangle}
\begin{document}

\title{ExtraFerm: An Extended Matchgate Simulator}

\author{
\IEEEauthorblockN{
Zack Hassman$^\dagger$,
Oliver Reardon-Smith$^\ddagger$,
Gokul Subramanian Ravi$^\ast$,
Frederic T. Chong$^\dagger$,
Kevin J. Sung$^\S$
}


\IEEEauthorblockA{
$^\dagger$\textit{University of Chicago}, USA\\
$^\ddagger$\textit{Center for Theoretical Physics, Polish Academy of Sciences}, Poland\\
$^\ast$\textit{University of Michigan}, USA \\
$^\S$\textit{IBM Quantum, IBM T.J. Watson Research Center}, USA
}

\IEEEauthorblockA{
zhassman@uchicago.edu, oreardonsmith@cft.edu.pl, gsravi@umich.edu, chong@cs.uchicago.edu, kevinsung@ibm.com}
}

\maketitle

\begin{abstract}
We present and open source Extraferm, a quantum circuit simulator tailored to chemistry applications. More specifically, our simulator can compute the Born-rule probabilities of samples obtained from circuits containing particle number-conserving matchgates and controlled-phase gates. We support both approximate and exact calculation of probabilities, and for approximate probability calculation, our simulator's runtime is exponential only in the magnitudes of the circuit's controlled-phase gate angles. This makes our simulator useful for simulating certain systems that are beyond the reach of conventional state vector methods. We demonstrate our simulator's utility by simulating the local cluster unitary Jastrow (LUCJ) ansatz and integrating it with sample-based quantum diagonalization (SQD) to improve the accuracy of molecular ground-state energy estimates with negligible computational overhead. More generally, we highlight a regime in which our simulator achieves substantially superior latency scaling and exponentially superior memory scaling over a tensor network simulator and a state vector simulator. As an efficient and flexible tool for simulating quantum chemistry circuits, our simulator enables new opportunities for enhancing near-term quantum algorithms in chemistry and related domains.
\end{abstract}

\begin{IEEEkeywords} matchgates, simulation, error mitigation, chemistry, sample-based quantum diagonalization \end{IEEEkeywords}

\section{Introduction}

Among the most promising near-term applications of quantum computing is quantum computational chemistry \cite{cao2019quantum, mcardle2020quantum}, which has attracted considerable attention due to potentially transformative applications in drug development \cite{santagati2024drug} and materials science \cite{alexeev2024quantum}, among other fields. Despite the excitement around this application, hardware demonstrations have been limited by the noise present in current quantum processors. Quantum error correction \cite{gottesman1997stabilizer, fowler2012surface, shor1995scheme, steane1996error, calderbank1996good, kitaev2003fault} offers a long-term solution to faulty hardware, but a fully fault-tolerant quantum computer has yet to be built. In the near-term, quantum error mitigation techniques~\cite{temme2017error,li2017efficient,huggins2021virtual,liao2024machine, cincio2021machine} can improve results obtained from noisy quantum processors, in which many of these techniques involve classical simulation of quantum circuits~\cite{czarnik2021error,filippov2023scalabletensornetworkerrormitigation,eddins2024lightcone}. Because classical simulation of quantum circuits is intractable in general, these simulations must introduce approximations, or are exact only when applied to specific narrow classes of circuits. Simulation methods tailored to circuits relevant to quantum chemistry may enable new techniques for enhancing the performance of chemistry workloads on quantum computers.

\begin{figure}[htbp]
    \centering
    \includegraphics[width=.45\textwidth]{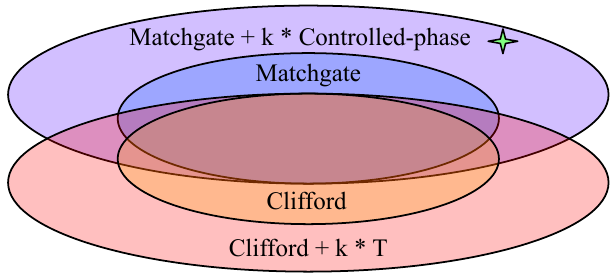}
    \caption{The green star indicates the class of circuits that ExtraFerm is designed to address: circuits containing $m$ matchgates and $k$ controlled-phase gates where $k \ll m$. This work concentrates on particle number-conserving circuits. Note that matchgates + controlled-phase gates form a universal gate set.}
    \label{fig:venn}
\end{figure}

In this work, we introduce ExtraFerm\footnote{https://github.com/zhassman/ExtraFerm}, an open-source quantum circuit simulator that can be used to compute the Born-rule probabilities of samples (bitstrings) from circuits composed of passive fermionic linear optical elements, also known as particle number-conserving matchgates, and controlled-phase gates. These circuits are highly relevant to the field of quantum computational chemistry for simulating quantum many-body systems. Fig.~\ref{fig:venn} shows where these circuits sit relative to other well-studied gate families. ExtraFerm can perform both exact and approximate probability calculations for measurements in the output distributions of these circuits. For approximate calculation, the only exponential component of ExtraFerm's runtime is a linear dependence in the \textit{extent} of the quantum circuit it is simulating. The extent is larger for controlled phase gates with larger phases and is multiplicative as more controlled phase gates are added (see Refs.~\cite{reardon2024improved,Dias2024classicalsimulation,reardonsmith2024fermioniclinearopticalextent}). For exact probability calculation, ExtraFerm's runtime is exponential only in the number of controlled-phase gates.

Notably, this means that for both approximate and exact calculation, ExtraFerm's performance is polynomial in the number of qubits and matchgates. Unlike conventional state vector simulation, which tracks the amplitudes of all bitstrings, ExtraFerm computes Born-rule probabilities for a pre-specified subset of the output distribution. Indeed, for small systems, one can use ExtraFerm to simulate the entire support of a circuit, but we do not expect this to be ExtraFerm's most compelling use case. Instead, we propose ExtraFerm as a unique tool that can be embedded in quantum-classical workflows to recover signal from noisy samples from large, application-scale quantum circuits.

We apply ExtraFerm to the simulation of the local unitary cluster Jastrow (LUCJ) ansatz~\cite{motta2023lucj}, which has recently been adopted for diverse applications in quantum simulation of chemical systems~\cite{blunt2025quantum, shajan2025toward, kaliakin2025implicit_solvent, liepuoniute2025quantumcentric, robledo2025sqd}. When mapped to a quantum circuit using the Jordan-Wigner transformation~\cite{ortiz2001quantum}, the LUCJ ansatz decomposes into particle number-conserving matchgates and controlled-phase gates, making it amenable to simulation by ExtraFerm in certain regimes. To highlight the regimes in which ExtraFerm may be useful, we use LUCJ circuits to study the error and latency performance of ExtraFerm. Note that LUCJ circuits are flexible in the sense that repeated layers of them can be used to implement arbitrary particle number-conserving extended matchgate circuits, providing us with a generic but practically useful benchmark. Beyond this, Section~\ref{comparison_existing} demonstrates a scenario in which ExtraFerm offers significantly lower latency and exponentially better memory performance compared to both a tensor network and state vector simulator.

Finally, as an end-to-end practical use case, we show how to use ExtraFerm to boost the performance of sample-based quantum diagonalization (SQD) \cite{robledo2025sqd}. SQD is an extension of quantum-selected configuration interaction (QSCI)~\cite{kanno2023quantum}, an algorithm that samples configurations from a quantum computer and uses them to select a subspace in which to diagonalize a molecular Hamiltonian. SQD adds an error mitigation procedure called configuration recovery that attempts to correct sampled bitstrings that were affected by noise and do not satisfy symmetries of the system. Configuration recovery greatly improves performance on noisy quantum processors, leading to the adoption of SQD for some of the largest demonstrations of quantum chemistry on quantum processors to date~\cite{liepuoniute2025quantumcentric,kaliakin2025implicit_solvent,shajan2025toward}.

We improve SQD by using ExtraFerm to select high-probability bitstrings during the early iterations of configuration recovery to provide better orbital occupancy information. We call this approach the ``warm-start'' variant of SQD and find that, compared to the original implementation of SQD, we obtain higher-accuracy energy estimates for 52-qubit and 60-qubit systems simulated on IBM Heron quantum processors. Furthermore, the overhead required to employ ExtraFerm and achieve these improvements is negligible. The main contributions of this work are the following:
\begin{enumerate}
    \item We release ExtraFerm, an extended matchgate simulator for particle number-conserving circuits allowing for high-performance calculation of Born-rule probabilities with tunable accuracy.
    \item We evaluate latency and error performance of ExtraFerm on local unitary cluster Jastrow (LUCJ) circuits to identify regimes for which ExtraFerm excels.
    \item We demonstrate how to integrate ExtraFerm with SQD, a hybrid quantum-classical energy estimation algorithm, to improve the accuracy of molecular ground-state energy calculations.
\end{enumerate}

\begin{figure*}[htbp]
    \centering
    \includegraphics[width=\textwidth]{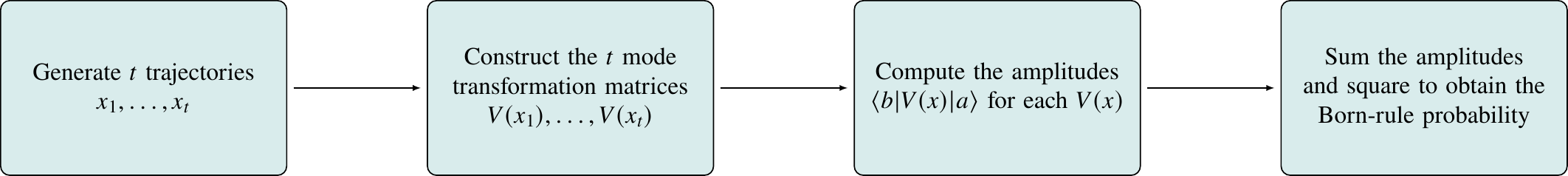}
    \caption{A high-level depiction of how ExtraFerm computes the Born-rule probability of a bitstring using \textsc{Raw Estimate} or \textsc{Exact}. The number of trajectories, $t$, can be given arbitrarily for \textsc{Raw Estimate}, while \textsc{Exact} sums over all possible trajectories. In practice, we parallelize this computation across bitstrings and then parallelize again across trajectories. Separate trajectories are generated for each bitstring to preserve the independence of Born-rule probability estimates.}
    \label{fig:simulator}
\end{figure*}

\section{Background}

\subsection{Classical Simulation of Quantum Circuits}

It is important to distinguish the very different computational tasks which are termed ``simulation''. ExtraFerm computes additive-precision estimates of individual Born-rule probabilities and can be thought of as an approximate strong simulator in the sense of Ref.~\cite{jozsa2013classical}. Conversely, a weak simulator would sample from the output distribution of a quantum circuit. While it might seem intuitive that a strong simulator is stronger than a weak simulator, the precise comparison between the different simulation tasks is subtle and depends on the error allowed~\cite{pashayan2020estimation}. However, our simulator can be used to approximately sample from the output distribution of a circuit when combined with the Qubit-by-qubit algorithm \cite{bravyi2022simulate} or the Gate-by-gate algorithm. 

An advantage of our methods is the ability to target specific Born-rule probabilities, while a state vector simulator computes all $2^n$ Born-rule probabilities for an $n$-qubit system. In applications such as the one we consider in Section~\ref{applications}, we are only interested in a small subset of the output distribution. As we will later discuss, the fact that we can use ExtraFerm to simulate arbitrary subsets of the output distribution allows us to improve SQD with very little computational overhead.

\subsection{The Theory of Matchgates}

Matchgates are a class of classically simulable quantum gates formalized by Valiant~\cite{valiant2002quantum}. Shortly afterwards, Terhal and DiVincenzo~\cite{terhal2002classical} discovered that matchgates are an equivalent expression of non-interacting fermions, giving these circuits a natural physical interpretation. The term `extended matchgate' refers to circuits that are primarily composed of matchgates but include a limited quantity of non-matchgates---in our case, controlled-phase gates. Together, matchgates and controlled-phase gates allow for universal quantum computation~\cite{PhysRevA.73.042313, brod2011extending}. However, most matchgate-based chemistry ansatze are particle number-conserving, meaning that the Hamming weights of all measurement outcomes are the same as the initial Hamming weight of the system, as specified by Pauli-X gates. These are the circuits that ExtraFerm is designed for.

More formally, a matchgate is a 2-qubit gate that has a $4 \times 4$ unitary matrix of the form 
\begin{equation}
G(A,B) = 
\begin{pmatrix}
    a_{11} & 0 & 0 & a_{12} \\
    0 & b_{11} & b_{12} & 0 \\
    0 & b_{21} & b_{22} & 0 \\
    a_{21} & 0 & 0 & a_{22}
\end{pmatrix}
\end{equation}
such that $\det(A) = \det(B)$ \cite{jozsa2008matchgates}.  Via the Jordan-Wigner transformation~\cite{ortiz2001quantum} we associate an $n$ qubit state with an $n$ mode fermionic state. A qubit state with qubit $i$ in state $\ket{1}$ maps to a fermionic state with a fermion in mode $i$. Let $M$ be a circuit of nearest-neighbor, particle number-conserving matchgates. Such a circuit preserves the Hamming weight of the computational basis state that it is applied to and is also sometimes called `passive' in the literature. The action of $M$
can be described by a unitary matrix $V$ which satisfies 
\begin{equation} 
Ma_i^\dagger M^\dagger = \sum_j V_{ij} a_j^\dagger
\end{equation}
where $a_i^\dagger$ denotes the $i^\text{th}$ creation operator defined via the Jordan-Wigner transformation. Note that while $M$ is a $2^n\times 2^n$ unitary matrix, $V$ is only $n\times n$. We refer to these `$V$' matrices as mode transformation matrices. Let $\ket{a}$ and $\ket{b}$ be two computational basis states with the same Hamming weights. Let $R$ be the indices of the bits of $\ket{b}$ that are equal to 1 and let $C$ be the indices of the bits of $\ket{a}$ that are equal to 1. Then
\begin{equation}
\bra{b}M \ket{a} = \det(\widetilde{V}).    
\end{equation}
where $\widetilde{V}$ is the submatrix of $V$ obtained by selecting rows $r \in R$ and $c \in C$ \cite{terhal2002classical}.

\subsection{Simulation of Extended Matchgate Circuits}


ExtraFerm builds upon the recently proposed mathematical framework from Reardon-Smith~\cite{reardon2024improved} and optimizes it for particle number-conserving circuits. We provide the details of this in section~\ref{simulator:algorithms}. Simultaneously with~\cite{reardon2024improved}, Diaz and Koenig~\cite{Dias2024classicalsimulation} developed alternative methods for extended matchgate simulation with essentially identical runtime scaling. Very recently, another approach has been developed with worse asymptotic scaling but significant improvements to the polynomial component of the runtime~\cite{wille2025classicalsimulationparitypreservingquantum}. These polynomial improvements may prove more relevant than the exponential component of the runtime in some parameter regimes. 

Finally, approaches involving decompositions of operators rather than state vectors have also been explored, e.g. in Refs~\cite{mocherla2023extending,Hakkaku_2022}, but these suffer from a substantially worse asymptotic scaling than state vector methods. For all of these methods, the runtime is dominated by an exponential sampling cost, but operator decomposition methods generally require a quadratic number of samples, i.e. the exponentially large component of the runtime is squared. Crucially, while there are many proposed mathematical frameworks for simulating extended matchgate circuits, there are few tools that exist for doing this in practice. ExtraFerm aims to contribute to filling this gap for particle number-conserving circuits.
\section{Simulator}
\subsection{Algorithms}
\label{simulator:algorithms}

ExtraFerm is capable of performing three different algorithms: \textsc{Raw Estimate}, \textsc{Estimate}, and \textsc{Exact}. \textsc{Raw Estimate} and \textsc{Estimate} are named after and follow the structure of those described in Ref.~\cite{pashayan2022fast}. These algorithms compute estimates of Born-rule probabilities, while \textsc{Exact} computes probabilities exactly. The \textsc{Raw Estimate} algorithm, which is depicted in Fig.~\ref{fig:simulator}, is the flagship capability of ExtraFerm. Given an additive error $\epsilon$, failure probability $\delta$, and probability upper bound $p_{\text{max}}$, \textsc{Raw Estimate} estimates the Born-rule probability $\hat{p}$ of a bitstring such that 
\begin{equation}
    \mathbb{P} \left[ \left| \hat{p} - p_{\text{true}} \right| > \epsilon \right] = 1 - \delta.
\end{equation}
\textsc{Raw Estimate} does this by summing up contributions from \textit{trajectories}, which are classically simulable matchgate approximations of the original circuit. Intuitively, each time we encounter a controlled-phase gate, we probabilistically branch between two possible paths. The number of trajectories that are needed to achieve a Born-rule probability estimate within $\epsilon$ for failure probability $\delta$ is dominated by the \textit{extent} $\xi^\ast$ of the circuit, given by the formula
\begin{equation}
    \xi^\ast = \prod_{j=1}^k \left( \cos \left(\frac{|\theta_j|}{4}\right) + \sin \left(\frac{|\theta_j|}{4}\right) \right)^2.
    \label{extent}
\end{equation}
where $\theta_j$ is the angle of the $j^{th}$ controlled-phase gate in the circuit. Using the parameters $\epsilon$, $\delta$, $p_{\text{max}}$, and $\xi^\ast$, a lower bound for the number of required trajectories is given by
\begin{equation}
    t = \left\lceil 2 \frac{\left( \sqrt{\xi^*} + \sqrt{p_{\max}} \right)^2}{\left( \sqrt{p_{\max} + \epsilon} - \sqrt{p_{\max}} \right)^2} \log \left( \frac{2e^2}{\delta} \right) \right\rceil
    \label{trajectories}.
\end{equation}

In practice, one may not always have particular error bounds in mind. To allow for heuristic use of ExtraFerm, users can directly provide an arbitrary trajectory count $t$ instead of $(\epsilon, \delta, p_{\text{max}})$. Furthermore, it is often not clear a priori what the associated $p_{\text{max}}$ is for a measurement outcome. Indeed, one can always provide $p_{\text{max}} = 1$, but this can lead to an unnecessarily large trajectory count, especially for small $\epsilon$ and $\delta$. In this case, one can use ExtraFerm' \textsc{Estimate} algorithm. This algorithm only requires $(\epsilon, \delta)$ as input parameters, and estimates Born-rule probabilities by starting with loose error and probability upper bounds followed by iterative calls to \textsc{Raw Estimate}. Finally, ExtraFerm can also compute exact probabilities with the \textsc{Exact} algorithm, which simply sums the contributions over all possible trajectories rather than just over the lower bound given by equation~\ref{trajectories}.

\subsection{Generating Trajectories}

\begin{figure*}[htbp]
    \centering
    \includegraphics[width=\textwidth]{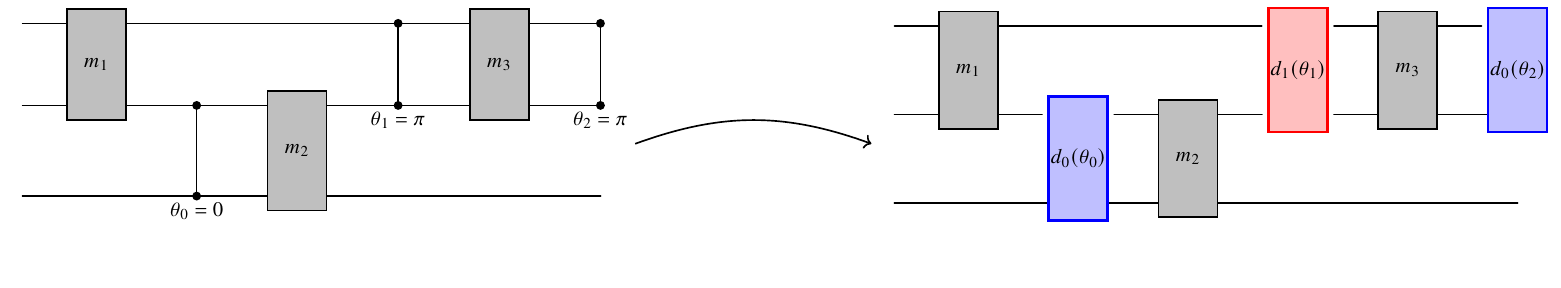}
    \caption{A visualization of how a trajectory $x$ and corresponding mode transformation matrix $V(x)$ are generated from a circuit. Each matchgate $m_i$ is unchanged while the controlled-phase gates are probabilistically assigned to either $d_0(\theta_j)$ or $d_1(\theta_j)$ based on $|\theta_j|$. In the above example, the trajectory $x = 010$ has been sampled. The first controlled-phase gate is assigned to $d_0(\theta_0)$ with probability $1$ while the second and third controlled-phase gates are assigned to $d_1(\theta_1)$ and $d_0(\theta_2)$, each with probability 1/2. Note that controlled-phase gates may operate on any two qubits; they are not restricted to nearest neighbors.}
    \label{fig:trajectory}
\end{figure*}

Consider a circuit composed of particle number-conserving matchgates and $k$ controlled-phase gates. Our ability to generate trajectories relies upon the following observation: notice that the $j^{th}$ controlled-phase gate $c(\theta_j )$ can be decomposed into a weighted sum of two unitary matrices, $d_0(\theta_j)$ and $d_1(\theta_j)$

\begin{align}
c(\theta_j)
&= e^{i\frac{\theta_j}{4}} \Bigg[
\cos\!\left(\tfrac{\theta_j}{4}\right)
\underset{d_0(\theta_j)}{\underbrace{
\begin{pmatrix}
    e^{-i\frac{\theta_j}{2}} & 0 & 0 & 0 \\
    0 & 1 & 0 & 0 \\
    0 & 0 & 1 & 0 \\
    0 & 0 & 0 & e^{i\frac{\theta_j}{2}}
\end{pmatrix}}} \nonumber \\
&\quad + i \sin\!\left(\tfrac{\theta_j}{4}\right)
\underset{d_1(\theta_j)}{\underbrace{
\begin{pmatrix}
    e^{-i\frac{\theta_j}{2}} & 0 & 0 & 0 \\
    0 & -1 & 0 & 0 \\
    0 & 0 & -1 & 0 \\
    0 & 0 & 0 & e^{i\frac{\theta_j}{2}}
\end{pmatrix}}}
\Bigg]
\end{align}
One can check that $d_0(\theta_j)$ and $d_1(\theta_j)$ are both particle number-conserving matchgates. For a circuit containing $k$ controlled-phase gates, there are $2^k$ ways to replace the controlled-phase gates with either $d_0(\theta_j)$ or $d_1(\theta_j)$. We
refer to these sequences as trajectories $x \in \{0,1\}^k$, for which the $j$th component $x_j$ indicates the choice for gate $d_{x_j}(\theta_j)$. We obtain the exact Born-rule probability $p_{\text{true}}$ with
\begin{equation}
    p_{\text{true}} = \left| \sum_{x \in \{0,1\}^k} w(x) \bra{b} V(x) \ket{a} \right|^2
\end{equation}
where
\begin{equation}
    w(x) = \prod_{j=1}^k \cos\left(\frac{\theta_j}{4}\right)^{x_j-1} \cdot \left(i \sin\left(\frac{\theta_j}{4}\right)\right)^{x_j}
\end{equation}
This calculation forms the basis of the \textsc{Exact} algorithm.

Allowing for a small additive error $\epsilon$ and failure probability $\delta$, Monte Carlo methods can be used to obtain an estimate for this Born-rule probability. Rather than summing up the amplitudes contributed by all $2^k$ trajectories, we can sample trajectories from the product probability distribution $P = \prod_{j=1}^k p(x_j)$:
\begin{align}
p(x_j = 0) &\equiv \frac{\cos\left(\frac{|\theta_j|}{4}\right)}{\sin\left(\frac{|\theta_j|}{4}\right) + \cos\left(\frac{|\theta_j|}{4}\right)} \\
p(x_j = 1) &\equiv \frac{\sin\left(\frac{|\theta_j|}{4}\right)}{\sin\left(\frac{|\theta_j|}{4}\right) + \cos\left(\frac{|\theta_j|}{4}\right)} 
\end{align}
\noindent To generate a trajectory, we use inverse transform sampling. For every controlled-phase gate angle $\theta_j$, we draw $u_j \sim \text{Unif}[0,1]$ and set
\begin{equation}
x_j =
\begin{cases}
1, & u_j < \dfrac{\sin\left(|\theta_j|/4\right)}
{\sin\left(|\theta_j|/4\right)+\cos\left(|\theta_j|/4\right)} \\
0, & \text{otherwise}
\end{cases}
\end{equation}

Observe that when $\theta_j = 0$, we select $d_0(\theta_j)$ with probability 1. Conversely, the worst case occurs when $\theta_j = \pi$ and we have an equal chance of selecting either $d_0(\theta_j)$ or $d_1(\theta_j)$. A visualization of this is provided by Fig.~\ref{fig:trajectory}.

\section{Implementation Details}

\subsection{Overview}

We have open sourced ExtraFerm on Github, and it can be installed with \texttt{pip install extraferm}. ExtraFerm consists of a Python interface and Rust backend. The Python interface integrates directly with Qiskit~\cite{qiskit}, while the Rust backend allows for multithreading of compiled binaries across multiple CPU cores, when available. Interaction between Python and Rust is facilitated by the PyO3 library, which allows Rust functions to be called directly from Python. 

The machinery behind ExtraFerm's \textsc{Raw Estimate} and \textsc{Exact} algorithms, depicted in Fig.~\ref{fig:simulator}, is doubly parallel. We use Rust's Rayon library to parallelize the per-bitstring probability computations as well as each bitstring's per-trajectory computations. A variety of other small optimizations led to further performance increases. For example, we represent states in a bit-packed format: Python uses native integers, while Rust leverages unsigned 128-bit integers (\texttt{u128}), enabling efficient bitwise operations. Additionally, we precompute and store values of $\sin(\theta_j/4)$ and $\cos(\theta_j/4)$ for all $\theta_j$, which reduces the number of computations inside hot loops.

\subsection{LUCJ-specific Optimizations}\label{lucj_optimizations}

Due to the growing popularity of the LUCJ ansatz, we equip ExtraFerm with specific optimizations for these circuits. LUCJ circuits, which we will denote with $\ket{\Psi}$ have the form
\begin{equation}
    \ket{\Psi} = e^{-\hat{K}_2} e^{\hat{K}_1} e^{i \hat{J}_1} e^{-\hat{K}_1} \ket{\mathbf{x_{\text{RHF}}}},
\end{equation}
for which $e^{-\hat{K}_2}, e^{\hat{K}_1}, e^{-\hat{K}_1}$ are orbital rotations, $e^{i \hat{J}_1}$ is a cluster operator, and $\ket{\mathbf{x_{\text{RHF}}}}$ is the restricted Hartree-Fock (RHF) state as described by the Jordan-Wigner transformation. The orbital rotation operators $e^{-\hat{K}_2}, e^{\hat{K}_1}, e^{-\hat{K}_1}$ of $\ket{\Psi}$ can be fully decomposed into particle number-conserving matchgates, while the operator $e^{i \hat{J}_1}$ can be fully decomposed into controlled-phase gates. 

Let $V_2$ be the mode transformation matrix describing the action of $e^{-\hat{K}_2}, e^{\hat{K}_1}$ and $V_1$ be the mode transformation matrix describing the action of $e^{-\hat{K}_1}$. Now let $D_{x_j}(\theta_j)$ be the mode transformation matrix representing the action of $d_{x_j}(\theta_j)$. Our circuit can be written with respect to a trajectory $x$ as
\begin{equation}
    V(x) = V_2 \left( \prod_{j=1}^k D_{x_j}(\theta_j) \right) V_1
\end{equation}
It is clear that $D_0^{-1}\cdot D_0 = I$, and one can check that
\begin{equation}
    D_0^{-1}(\theta_j) \cdot D_1(\theta_j) 
    = \text{Diag}\bigl(1, \ldots, 
    \underbrace{-1}_{q1}, \ldots, 
    \underbrace{-1}_{q2}, \ldots, 1\bigr)
\end{equation}
where $q_1, q_2$ are the two qubits that $D_{x_j}(\theta_j)$ acts on. Notice that $W \equiv \prod_{j=1}^k D_0^{-1}(\theta_j) \cdot D_{x_j}(\theta_j)$ is a diagonal matrix with only $\{+1, -1\}$ on its diagonal. Notably, $W$ has no dependence on $\{\theta_j\}$. Define $V_3 = V_2 \cdot \prod_{j=1}^k D_0(\theta_j)$ and reformulate $V(x)$ as
\begin{align}
    V(x) &= V_2 \left( \prod_{j=1}^k D_0(\theta_j) \cdot D_0^{-1}(\theta_j) \cdot D_{x_j}(\theta_j) \right) V_1 \\
    &= V_3 \left( \prod_{j=1}^k D_0^{-1}(\theta_j) \cdot D_{x_j}(\theta_j) \right) V_1 \\
    &= V_3 \left(I - 2 \cdot \sum_{i \in N} E_{ii} \right) V_1 \\
    &= V_3 V_1 - 2 \cdot \sum_{i \in N} V_3 E_{ii} V_1
\end{align}
where $N = \{i | W_{ii} = -1\}$ and $E_{ii}$ is a standard basis matrix. In this sense, any possible $V(x)$ can be written as the sum of $V_3V_1$ and a small number of corrections of the form $V_3 E_{ii} V_1$ applied to it. ExtraFerm computes all of these matrices once at runtime. Additionally, ExtraFerm stores the determinants of each of these unique patterns in a hash table to avoid redundant determinant evaluations across trajectories. 

These optimizations greatly reduce the two primary costs of Born-rule probability computation, namely, the construction of mode transformation matrices $V(x)$ and the evaluation of their determinants. We find that these optimizations lead to speedups of up to two orders of magnitude for LUCJ circuits over our standard implementation of \textsc{Raw Estimate}. When ExtraFerm detects an LUCJ circuit, this optimized path is automatically taken. 

\begin{figure}[htbp]
    \centering
    \includegraphics[width=0.48\textwidth]{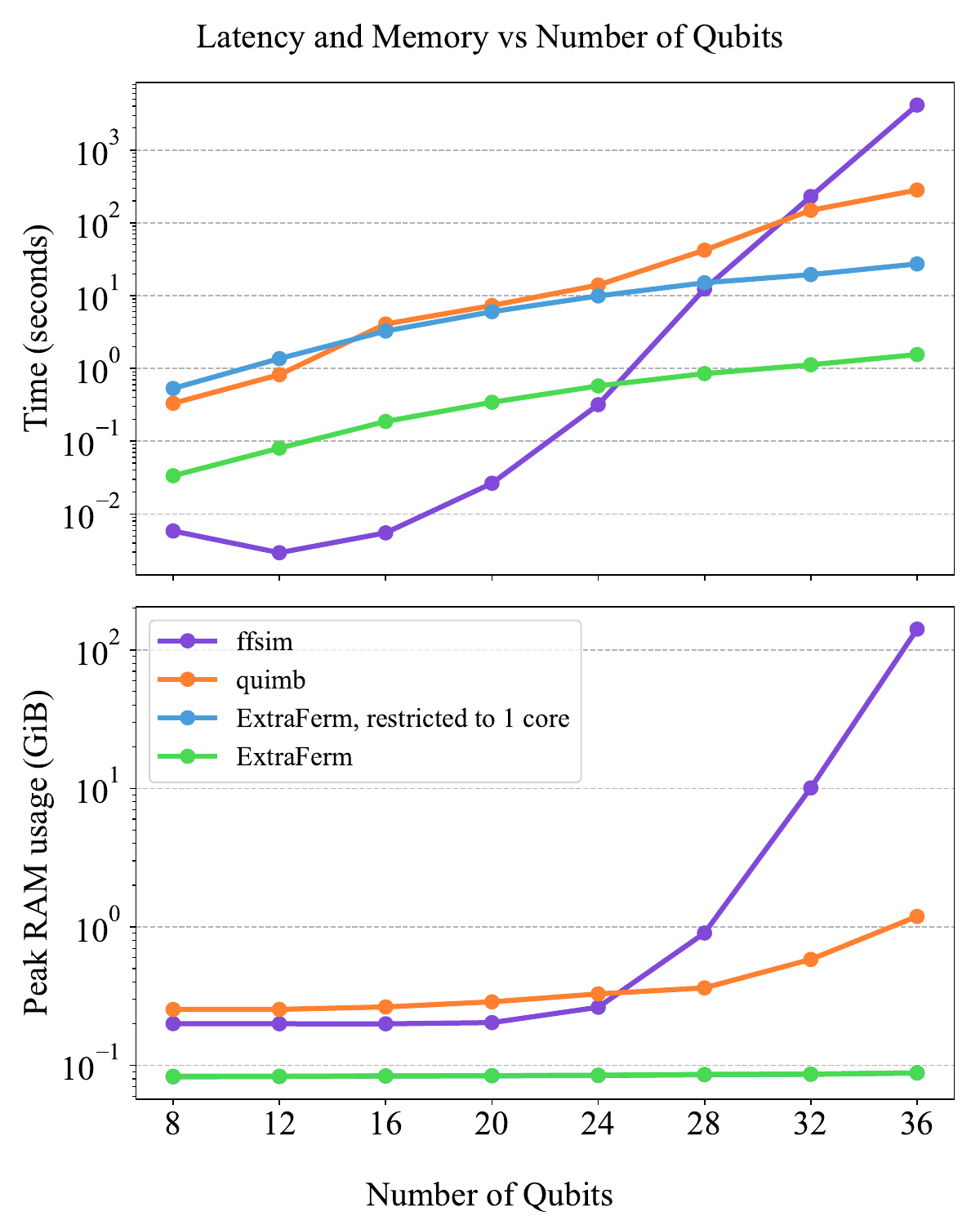}
    \caption{Latency and memory comparisons of different tools when used to calculate the exact probability of an outcome measurement from a randomly generated particle number-conserving extended matchgate circuit. Each circuit contains 16 controlled-phase gates and a quantity of matchgates proportional to circuit size. Results are collected on an Intel Xeon w7-2495X processor. By default, all tools are given access to 24 cores, but only ExtraFerm benefits significantly from using more than 1. We also include results for ExtraFerm restricted to 1 core. Note that the y-axis for each graph is on the logarithmic scale (higher values are orders of magnitude worse).}
    \label{fig:latency_memory}
\end{figure}

\section{Benchmarking and Error Analysis}

\subsection{Comparisons with Existing Tools}
\label{comparison_existing}

We emphasize that while LUCJ circuits are an important use case for ExtraFerm, our simulator is a competitive tool for simulating particle number-conserving extended matchgate circuits in general. To demonstrate this, we compared ExtraFerm's latency and memory consumption against quimb~\cite{gray2018quimb}, a tensor network simulator and ffsim~\cite{ffsim}, a state vector simulator that takes advantage of particle number symmetry. While ffsim computes probabilities for the entire output distribution, quimb is similar to ExtraFerm in the sense that it can compute the probability of a single outcome measurement. We benchmark these three tools on randomly-initialized particle number-conserving extended matchgate circuits ranging from 8 to 36 qubits. These circuits contain 16 controlled-phase gates and a quantity of matchgates proportional to the circuit's size. Note that these more general circuits are not amenable to the optimizations for LUCJ circuits described in Section \ref{lucj_optimizations} and thus these optimizations are not applied.

Fig.~\ref{fig:latency_memory} demonstrates that for ExtraFerm, there is a substantially superior latency scaling over quimb and ffsim. ExtraFerm's embarrassingly parallel implementation allows it to take full advantage of all 24 CPU cores on the machine that we perform this benchmark on. On the other hand, quimb and ffsim do not significantly benefit from using more than one core. ExtraFerm's latency scaling advantage for large circuits is still substantial but less extreme when ExtraFerm is restricted to a single CPU core. In terms of memory consumption, system size hardly has an effect on ExtraFerm, while quimb and ffsim suffer an exponential blow-up beginning around the 28-qubit mark.

We note that while ffsim is likely to be state-of-the-art among state vector simulators, our usage of quimb can probably be improved on. For simplicity, we used its exact tensor network contraction functionality, but it might be possible to introduce truncations that leave the result unaffected up to numerical precision. While we placed the controlled-phase gates between random pairs of qubits for our benchmarks, recent work has shown that tensor networks can be effective at simulating certain LUCJ circuits with controlled-phase gates that are geometrically local on a 2-dimensional lattice~\cite{rudolph2025simulating}.

\begin{figure}[htbp]
    \centering
    \includegraphics[width=0.48\textwidth]{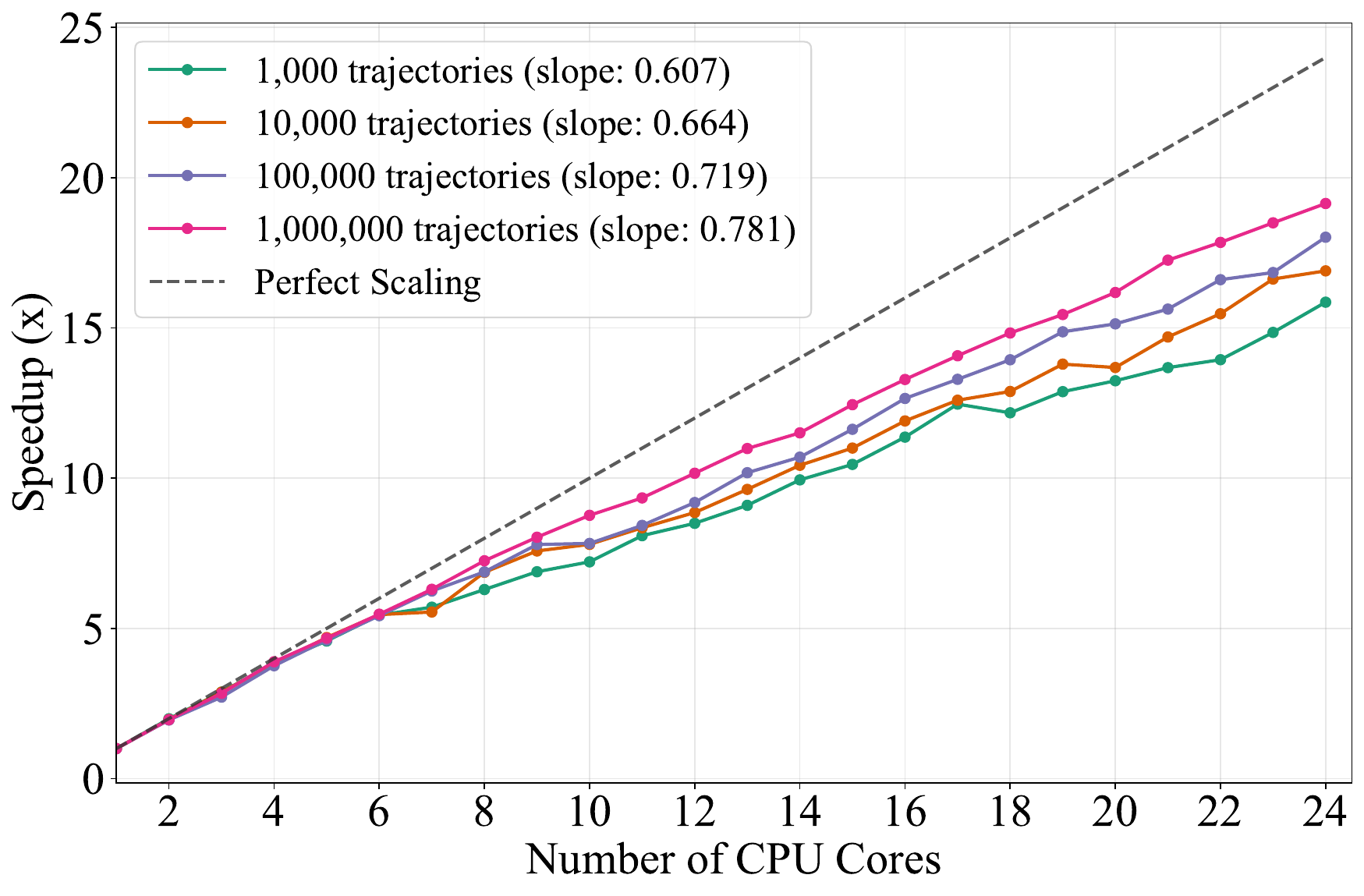}
    \caption{Speedup as a function of the number of cores used on a 24-core Intel Xeon w7-2495X processor for executing ExtraFerm on a 40-qubit LUCJ circuit containing 43 controlled-phase gates with angles sampled from $\mathcal{N}(0, 10^{-3})$ and a total extent 1.732. For each data point, \textsc{Raw Estimate} was used to compute probabilities for 1,000 bitstrings and times were averaged over 10 trials.}
    \label{fig:speedup_cores}
\end{figure}

\subsection{Latency Benchmarks}

In this section, we focus on the \textsc{Raw Estimate} algorithm, the fundamental subroutine of the \textsc{Estimate} and an approximate version of the \textsc{Exact} algorithm. \textsc{Raw Estimate} can be run with an arbitrary number of trajectories, making it more flexible than \textsc{Exact}. At its core, a \textsc{Raw Estimate} computation amounts to summing contributions from a multiset of sampled trajectories. Each trajectory corresponds to a particular decomposition of the controlled-phase gates and contributes an expectation value that must be accumulated into the overall estimate. Accordingly, we examine its runtime behavior, with attention not only to the number of trajectories required but also to non-asymptotic considerations that significantly impact the cost of evaluating each trajectory.

\begin{figure}[htbp]
    \centering
    \includegraphics[width=0.48\textwidth]{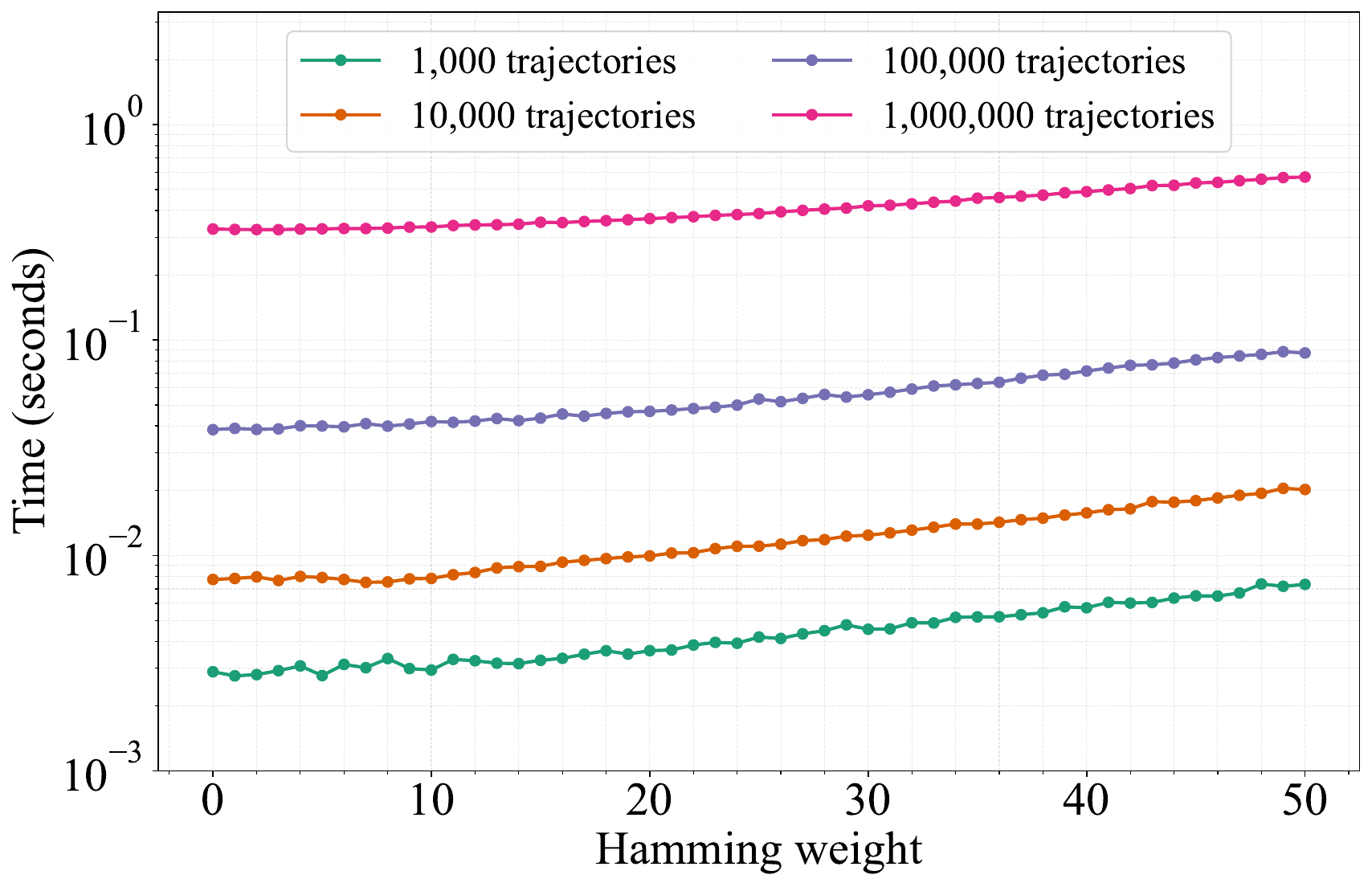}
    \caption{The average runtime required to estimate a bitstring probability using \textsc{Raw Estimate} for a 100-qubit LUCJ circuit containing 111 controlled-phase gates with gate angles drawn from $\mathcal{N}(0,10^{-3})$ and a total circuit extent of 5.281. Data points were collected for Hamming weights (the number of fermions) ranging from 0 to 100 in increments of 2 and averaged over 10 trials. Simulations were performed on a 24-core Intel Xeon w7-2495X processor.}
    \label{fig:time_vs_hamming}
\end{figure}

For a given trajectory $x$, the two main costs are constructing the mode transformation matrix $V(x)$ and evaluating the expectation value $\bra{b} V(x) \ket{a}$. The latter is dominated by determinant calculations, which require selecting an $h \times h$ submatrix of $V(x)$, where $h$ is the Hamming weight of $\ket{a}$ and $\ket{b}$. In the computational basis, $h$ corresponds to the number of set bits, or equivalently, the number of $X$ gates applied to prepare the Hartree-Fock state in the context of LUCJ circuits. Thus, states with larger Hamming weights---i.e., circuits with more fermions---incur higher computational cost. This can be seen clearly in Fig.~\ref{fig:time_vs_hamming}, in which we fix the angles of a 100-qubit LUCJ circuit and compare the total time required to estimate the probability of a bitstring as a function of its Hamming weight. Independent of the number of trajectories, there is a clear trend indicating higher execution times associated with larger Hamming weights. In addition, Fig.~\ref{fig:time_vs_hamming} shows that for any given Hamming weight, when the number of trajectories is increased by a factor of 10, the respective runtime is also increased by approximately a factor of 10, as we would expect.

However, regardless of the Hamming weight of our bitstring, increasing the number of cores available to ExtraFerm yields near-linear speedups. This is demonstrated in Fig.~\ref{fig:speedup_cores}, which shows execution times for running \textsc{Raw Estimate} to estimate the probabilities of 1,000 bitstrings in parallel for different quantities of trajectories. Furthermore, we can see that as the task becomes harder (more trajectories are used), ExtraFerm scales more effectively.

\subsection{Error Analysis}

\begin{figure*}[htbp]
    \centering
    \includegraphics[width=\textwidth]{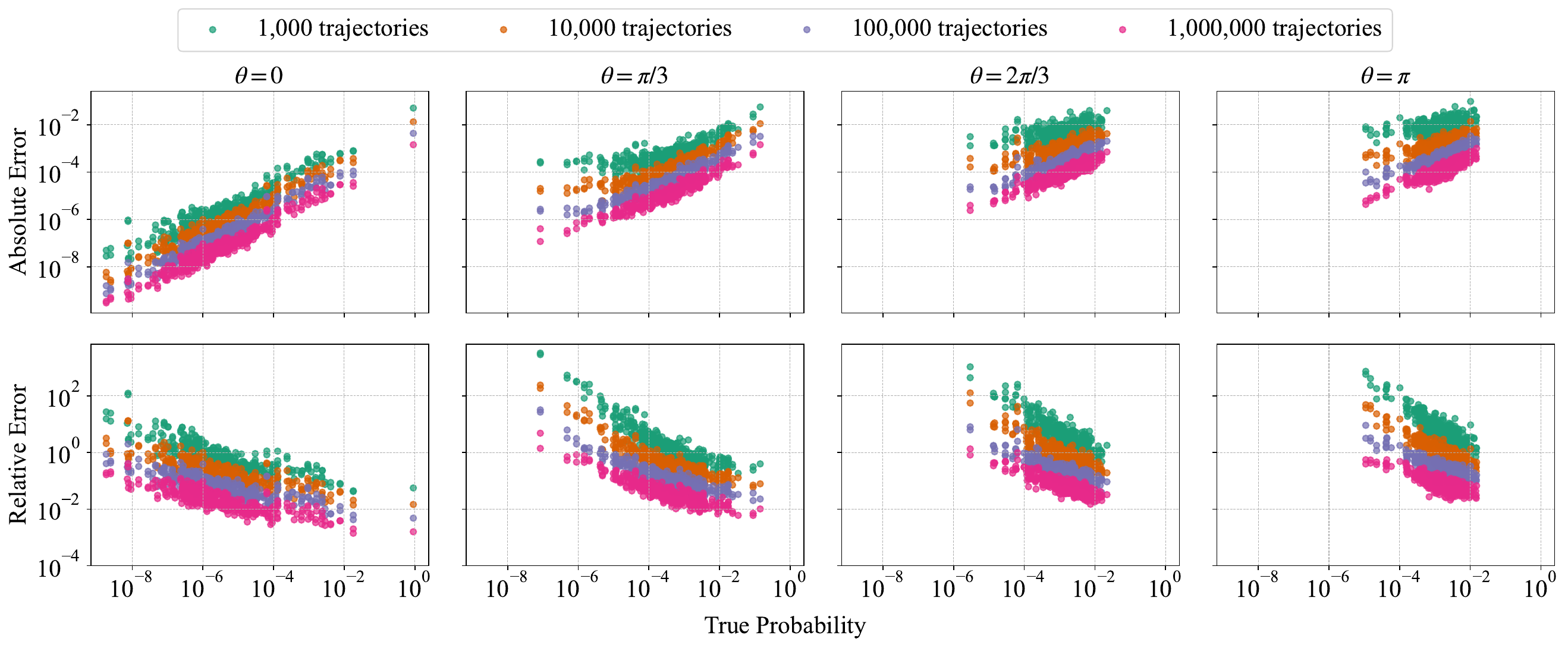}
    \caption{The error between the true probability and probability obtained from \textsc{Raw Estimate} for all 400 outcome measurements in the support of a 12-qubit LUCJ circuit with 12 controlled-phase gates. These circuit correspond to fermionic systems with 6 spatial orbitals, 3 alpha spin electrons, and 3 beta spin electrons. They have extents of 3.753, 228.091, 2,254.514, 3,862.055, respectively (left to right). Orbital rotations are randomly initialized while controlled phase gate angles are sampled i.i.d. from $\sim \mathcal{N}(\theta, 0.1)$. Results are displayed for $\theta \in \{0, \pi/3, 2\pi/3, \pi\}$ across trajectory counts of 1,000, 10,000, 100,000, and 1,000,000. Each point is averaged over 10 trials. The top row displays absolute error $|\hat{p} - p_{\text{true}}|$ as a function of $p_{\text{true}}$ while the bottom row displays relative error $|\hat{p} - p_{\text{true}}| / p_{\text{true}}$ as a function of $p_{\text{true}}$. }
    \label{fig:error_vs_probability}
\end{figure*}

\begin{figure*}[htbp]
    \centering
    \includegraphics[width=\textwidth]{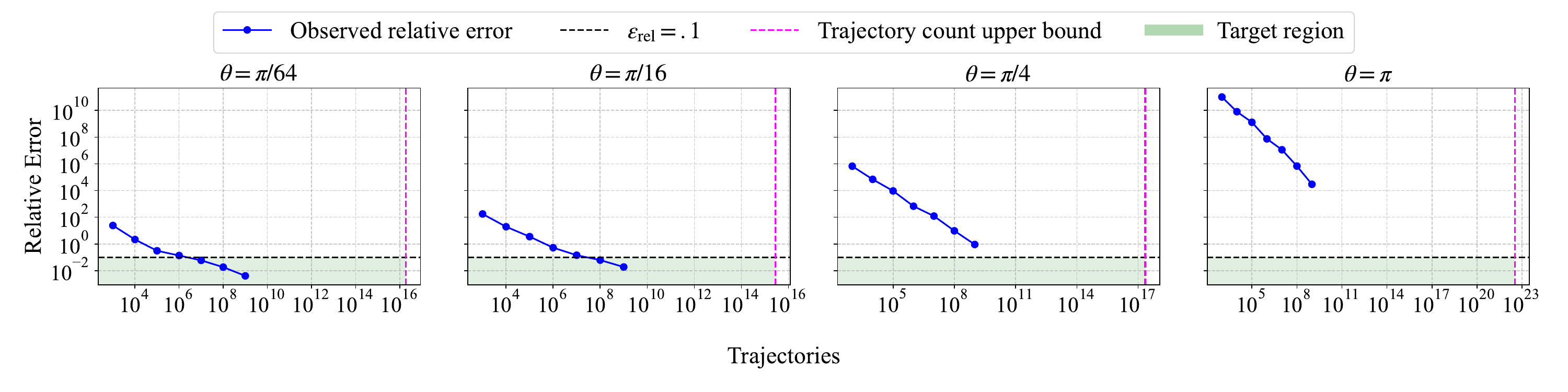}
    \caption{Relative error of \textsc{Raw Estimate} versus trajectory counts for a 40-qubit LUCJ circuit with 43 controlled-phase gates. These circuits correspond to fermionic systems with 20 spatial orbitals, 2 alpha spin electrons, and 2 beta spin electrons. They have extents of 3.48, 62.0, $1.22 \cdot 10^6$, and $8.77 \cdot 10^{12}$, respectively (left to right). Orbital rotations are randomly initialized while controlled phase gate angles are sampled i.i.d. from $\mathcal{N}(\theta,10^{-3})$ with $\theta\in \{\pi/64,\pi/16,\pi/4,\pi \}$. Each plot uses the median-probability bitstring from its respective circuit, so the bitstring and true probability $p_{\text{true}}$ differ across panels. Blue markers show the mean relative error over 10 trials. The horizontal black dashed line marks relative error $\epsilon_{\text{rel}}=.1$ while the vertical magenta dashed line is the upper bound on the number of trajectories needed to achieve absolute error $\epsilon_{\text{abs}}=p_{\text{true}} \cdot .1$ with $\delta=10^{-4}$ (using $p_{\max}=p_{\text{true}}$). The green shading highlights the region where both conditions hold: $|\hat p-p_{\text{true}}|\le p_{\text{true}} \cdot .1$ and the trajectory count is at or below the upper bound.}
    \label{fig:rel_error_vs_trajectories}
\end{figure*}

We now share error trends observed when using \textsc{Raw Estimate} to estimate probabilities for LUCJ circuits. We begin with a case study of a 12-qubit LUCJ circuit for which we estimated the probabilities of all outcome measurements in the support with different trajectory counts while varying the average controlled-phase gate angles. In Fig.~\ref{fig:error_vs_probability}, we display both the observed absolute and relative error for the entire support of the circuit. Across all circuits, we observe that absolute error is proportional to the magnitudes of Born-rule probabilities while relative error is inversely proportional to the magnitudes of Born-rule probabilities. The fact that a bitstring's absolute error is proportional to its Born-rule probability ends up being very useful in practice. In section \ref{applications}, we use this to successfully distinguish high-probability bitstrings from low-probability bitstrings with very small computational overhead.

For all four circuits, we observe decreases in both absolute and relative error as the trajectory count increases. This trend is consistent with the theoretical guarantees of \textsc{Raw Estimate}, under which additive error decreases as trajectory count grows. Comparisons between the four circuits are more nuanced. Varying the controlled-phase gate angles alters the Born-rule probabilities of the measurement outcomes. As these angles increase, the output distribution flattens: probability mass spreads across a larger fraction of the support, reducing the dominance of a few high-probability outcomes. This redistribution generally increases both absolute and relative errors. Thus, while increasing trajectory counts consistently reduces error across all circuits, larger controlled-phase angles shift the overall error profile upward.

This phenomenon is illustrated in Fig.~\ref{fig:rel_error_vs_trajectories}, which reports the relative error as a function of trajectory count. We examine four 40-qubit LUCJ circuits that again differ only in the angles of their controlled-phase gates. To isolate the effect of these angles on error behavior, we select the median-probability bitstring from the support of each circuit and track its estimated error across increasing trajectory counts. The specific bitstring differs between circuits, but in each case we mark on the $x$-axis the theoretical trajectory count required to achieve a relative error of $0.1$, with the same threshold also indicated by a dashed line on the $y$-axis. For small controlled-phase gate angles ($\theta \approx 0$), this target is reached with relatively few trajectories. As $\theta$ increases, however, the observed errors remain greater than the desired level, and the error curves become steeper, demonstrating a larger difference in performance based on the number of trajectories used.
\section{Applications}\label{applications}

\begin{figure*}[htbp]
    \centering
    \includegraphics[width=\textwidth]{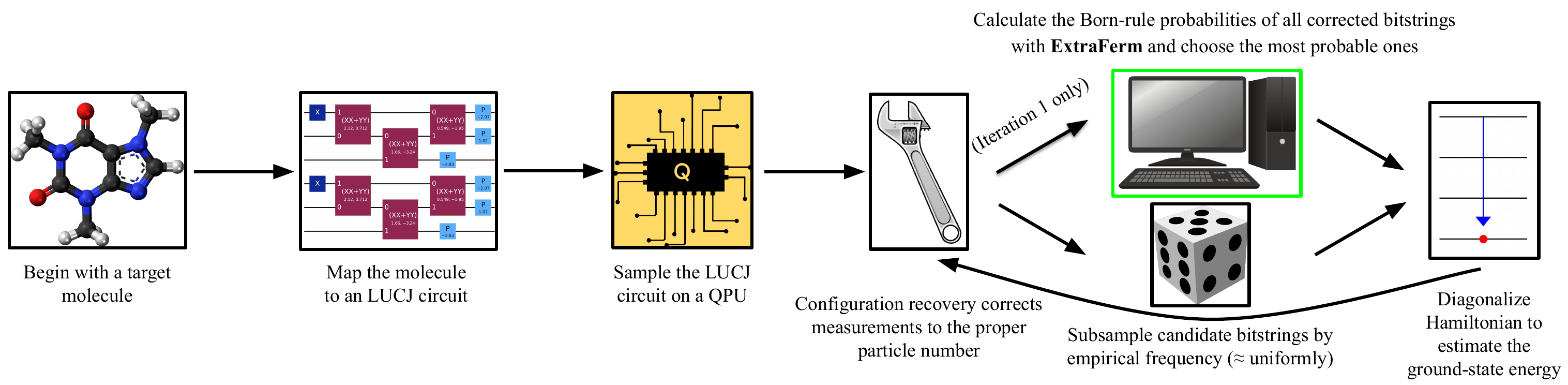}
    \caption{A schematic of our modification to the SQD algorithm. For the warm-start approach, the top path is taken after the first iteration of configuration recovery, and ExtraFerm is used to calculate the Born-rule probabilities of bitstrings that have been corrected to the proper particle number. The gray loop below is performed for the subsequent iterations of SQD. Our experiments demonstrate that this modification can improve the accuracy of ground-state energy calculations.}
    \label{fig:workflow}
\end{figure*}

\subsection{Sample-Based Quantum Diagonalization}
Sample-Based Quantum Diagonalization (SQD) is an algorithm introduced in Ref.~\cite{robledo2025sqd} as an extension of Quantum-Selected Configuration Interaction \cite{kanno2023quantum}. The central idea is to sample chemistry circuits on noisy quantum hardware and then apply classical post-processing to estimate molecular ground-state energies. A key property of LUCJ circuits is that, beyond conserving the total Hamming weight of their initial states, they also preserve the Hamming weights of the $\alpha$- and $\beta$-spin sectors independently. Equivalently, an LUCJ state can be expressed as $\ket{\Psi} = \ket{\beta \, \alpha}$, where the spin sectors must retain their individual Hamming weights after the circuit is applied. SQD corrects bitstrings measured on quantum hardware that violate these conservation rules prior to using them in the energy calculation. Note that SQD is a variational algorithm, which guarantees that the energy estimates produced by it can never be lower than the ground-state energy. In other words, lower energy estimates are better.

We give a brief overview of SQD. First, a molecule is mapped to an LUCJ circuit $\ket{\Psi}$. Second, this circuit is sampled on a quantum computer to collect $\widetilde{\mathcal{X}}$, a multiset of noisy bitstrings representing electronic configurations. Third, bitstrings with incorrect particle number are then restored through an iterative procedure called \textit{configuration recovery} to produce the multiset of bitstrings $\mathcal{X}_\text{R} = \mathcal{X}_{N} \cup \mathcal{X}_{\rightarrow N}$ where $\mathcal{X}_{N}$ denotes the bitstrings that originally had the correct particle number and $\mathcal{X}_{\rightarrow N}$ denotes the bitstrings that have been restored to the correct particle number. Fourth, $\mathcal{S}$ is subsampled from $\mathcal{X}_\text{R}$ according to observed bitstring frequencies, and the chemistry Hamiltonian is diagonalized in this subspace to obtain an energy value. Lastly, the average orbital occupancies are calculated, which are used to inform the Hamiltonian diagonalization step during the next iteration of SQD.

\subsection{Warm-Start Approach}

We propose a simple, cost-effective modification to the SQD workflow that we call \textit{warm-starting} (see Fig.~\ref{fig:workflow}). For the original SQD workflow, $\mathcal{S}$ is subsampled from  $\mathcal{X}_\text{R}$ according to the empirical frequencies of the bitstrings in  $\mathcal{X}_\text{R}$. However, due to the large support of these circuits, each corrected bitstring in $\mathcal{X}_\text{R}$ appears extremely infrequently for nontrivial chemical systems. This means that $\mathcal{S}$ is subsampled approximately uniformly from $\mathcal{X}_\text{R}$, that is, each bitstring is chosen with a probability of $\approx \frac{1}{\# \text{ shots}}$. Since SQD is a variational algorithm, this method of subsampling works well, allowing the solver to explore a diverse range of electronic configurations and converge to the ground state energy.

However, because average orbital occupancies identified during early iterations of SQD inform the Hamiltonian diagonalization step during later iterations of SQD, energy estimates can suffer from high variance between trials. Our warm-start approach addresses this by using ExtraFerm to calculate the probabilities of the bitstrings in $\mathcal{X}_\text{R}$ during the first iteration of configuration recovery. We then choose the bitstrings with the highest probabilities, rather than subsampling. Later iterations of SQD proceed as usual, subsampling from $\mathcal{X}_\text{R}$ according to the empirical frequencies of the corrected bitstrings. We find that the warm-start approach can produce energy estimates that are on average more accurate than the ones produced by the original SQD workflow. 

\subsection{Experimental Setup}

We present energy curves for two systems: 1) \ce{N_2}, molecular nitrogen in the cc-pVDZ basis set and  2) \ce{H_{30}}, a hydrogen chain in the STO-6G basis set. For a given bond distance, each system was mapped to an LUCJ circuit initialized using CCSD parameters, following Ref.~\cite{robledo2025sqd}. We sampled 100,000 shots from each circuit on a Heron QPU and then performed Sample-Based Quantum Diagonalization (SQD) for 5 iterations and with 1 batch. By construction, SQD performs diagonalization in a subspace with dimension equal to the square of the batch size, so the subspace dimension is always a perfect square. To demonstrate the scaling of these calculations, we collected results for a range of subspace dimensions: $\{500^2, 1000^2, 1500^2, 2000^2\}$. For the warm-start variant of SQD, Extraferm computed probabilities with the \textsc{Raw Estimate} algorithm using 1,000 trajectories per bitstring. Across each combination of method, dimension, and bond distance, the SQD calculations were performed on 12 cores of an Intel Gold 6248R processor and averaged over 5 trials. QPU samples were reused across trials.

The energy curve for the nitrogen molecule \ce{N2} includes 49 bond distances, evenly spaced between $0.70000-3.1000$ Å. Each corresponding 52-qubit circuit contained 2,104 matchgates and 57 controlled-phase gates, with extents ranging from 1.129 to 3.281 with an average extent of 2.026. These were sampled on the IBM Boston QPU. These circuits correspond to fermionic systems with 26 spatial orbitals, 5 alpha spin electrons, and 5 beta spin electrons with a total support of $\binom{27}{5}^2 = 6.517 \times 10^9$ bitstrings.

The energy curve for the hydrogen chain \ce{H30} includes 15 bond distances, evenly spaced between $0.79377-1.16419$ Å. Each 60-qubit circuit contained 2,790 matchgates and 66 controlled-phase gates, with extents ranging from 1.061 to 1.662 with an average extent of 1.259. These were sampled on the IBM Torino QPU. These circuits correspond to fermionic systems with 30 spatial orbitals, 15 alpha spin electrons, and 15 beta spin electrons with a total support of $\binom{30}{15}^2 = 2.406 \times 10^{16}$ bitstrings, significantly larger than that of the \ce{N2} system.

\begin{figure*}[htbp]
    \centering
    \includegraphics[width=\textwidth]{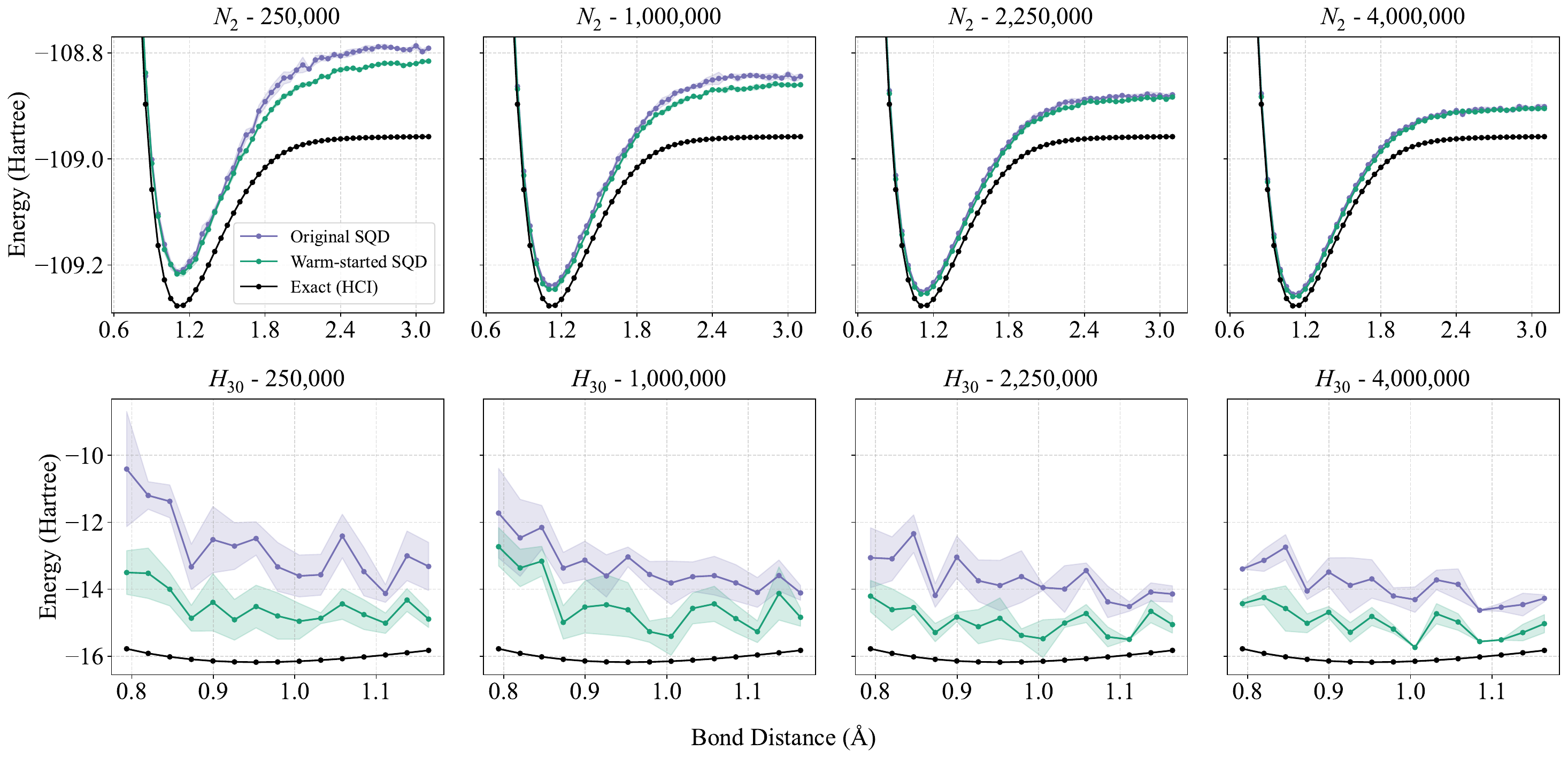}
    \caption{A comparison of SQD and warm-started SQD for \ce{N2} and \ce{H30}. For all plots, the shadows show $\pm 1$ standard deviation across 5 trials. \textbf{Top row}: ground-state energy estimates for \ce{N2} at 49 bond distances (0.70000–3.10000 Å) using 52-qubit LUCJ circuits. \textbf{Bottom row}: ground-state energy estimates for \ce{H30} at 15 bond distances (0.79377–1.16419 Å) using 60-qubit LUCJ circuits.}
    \label{fig:energies}
\end{figure*}

As shown in Fig.~\ref{fig:energies} and Table~\ref{tab:merged}, the reductions in error when using the warm-starting technique ranged between $13.31-18.71 \%$ for the nitrogen system and $43.26-54.89 \%$ for the hydrogen system relative to the heat-bath configuration interaction (HCI) energy. The greater improvement for hydrogen may be attributed to the larger system size and higher initial error. We report SQD runtime as the average time required to perform configuration recovery, run ExtraFerm (if warm-starting), and diagonalize the Hamiltonian. ExtraFerm's overhead---reported as the added percentage of the tool's runtime to the SQD runtime---was negligible, accounting for at worst an additional $1.98 \%$. While these energy estimates are not within the 1.6 milliHartree error threshold (an often-used precision target), ExtraFerm and the warm-starting technique constitute another step towards quantum utility for chemistry simulations.

\begin{table*}[htbp]
\centering
\caption{Summary of Results for \ce{N2} and \ce{H30} Systems}
\label{tab:merged}

\begin{tabular}{l c c c c c}
\multicolumn{6}{c}{\textbf{\ce{N2}}}\\[0.3em]
\hline
\textbf{Subspace Dim.} & \textbf{Avg. Original Error} & \textbf{Avg. Warm Error}  & \textbf{Avg. Error Red.} & \textbf{SQD Runtime per Bond Dist.} & \textbf{ExtraFerm Overhead} \\     
\hline
$500^2 = 250,000$ & 0.109 \% & 0.090 \% & 14.87 \% & 3.15 min & 1.98 \% \\
$1,000^2 = 1,000,000$ & 0.071 \% & 0.058 \% & 18.71 \% & 6.09 min & 1.02 \% \\
$1,500^2 = 2,250,000$ & 0.049 \% & 0.042 \% & 15.14 \% & 9.79 min & 0.63 \% \\
$2000^2 = 4,000,000$ & 0.035 \% & 0.031 \% & 13.31 \% & 14.59 min & 0.42 \% \\
\hline
\\[-0.3em]

\multicolumn{6}{c}{\textbf{\ce{H30}}}\\[0.3em]
\hline
\textbf{Subspace Dim.} & \textbf{Avg. Original Error} & \textbf{Avg. Warm Error}  & \textbf{Avg. Error Red.} & \textbf{SQD Runtime per Bond Dist.} & \textbf{ExtraFerm Overhead} \\    
\hline
$500^2 = 250,000$ & 20.66 \% & 9.47 \% & 53.81 \% & 0.62 h & 0.43 \% \\
$1,000^2 = 1,000,000$ & 16.98 \% & 9.94 \% & 43.26 \% & 2.97 h & 0.09 \% \\
$1,500^2 = 2,250,000$ & 14.56 \% & 6.58 \% & 54.89 \% & 6.12 h & 0.04 \% \\
$2000^2 = 4,000,000$ & 13.36 \% & 6.41 \% & 53.41 \% & 10.91 h & 0.02 \% \\
\hline
\end{tabular}

\end{table*}

\section{Related Works}

\subsection{Simulation of Clifford+T Circuits}

The Clifford+T gate set, comprising the Clifford group (generated by the phase, Hadamard, and CNOT gates) and the T gate, is among the most extensively studied in quantum computing. Stabilizer circuits, which contain only Clifford gates, are classically simulable in polynomial time by the Gottesman-Knill theorem~\cite{gottesman1998heisenberg}. Adding T gates enables universal quantum computation~\cite{bravyi2005universal}. States produced by a small number of T gates have low stabilizer rank and can be expressed as a concise linear combination of stabilizer states~\cite{bravyi2019simulation}, making them amenable to classical simulation. Bravyi and Gosset~\cite{bravyi2016improved} leverage this property to introduce a Monte Carlo probability estimation routine for Clifford-dominated circuits. This algorithm is similar in spirit to Reardon-Smith's~\cite{reardon2024improved} (which ExtraFerm builds upon), though it is tailored to a different class of circuits.

\subsection{Other Simulation Frameworks}

Tensor networks \cite{white1992density, orus2014practical} are a framework for representing quantum many-body systems and are a popular choice for simulating quantum circuits \cite{markov2008simulating}. They express quantum states as a collection of complex-valued multidimensional arrays. However, tensor network methods may scale poorly for systems with high entanglement \cite{berezutskii2025tensor}. Recent work has simulated the LUCJ ansatz using tensor networks~\cite{rudolph2025simulating}. Using a different approach for this circuit family, Ref.~
\cite{belagali2026efficientclassicalsimulationlargescale} developed a classical algorithm to calculate the expectation value of a one-layer unitary cluster Jastrow circuit in polynomial time.

Alternatively, Pauli propagation~\cite{rudolph2025paulipropagationcomputationalframework} methods approximate the evolution of a quantum operator by truncating the Pauli path integral, enabling fast, rough estimates of expectation values even on arbitrary circuit topologies. Initially shown to be effective in noisy and random settings, the approach has since been applied more broadly and appears promising for variational use cases. Pauli propagation simulation costs scale exponentially with non-Clifford gates and accuracy requirements.

\subsection{Classical Simulation to Boost Quantum Fidelity}

Recently, several works have used classical simulation to mitigate errors on quantum computers. These methods typically simulate a subset of the quantum system or an efficiently simulable related system. For instance, Clifford Data Regression~\cite{czarnik2021error} fits a linear model using Clifford-dominated circuits to map noisy observables to their noiseless counterparts, achieving error reductions between one and two orders of magnitude. Ref.~\cite{liu2022classical} offloads the computation of an auxiliary error-mitigation circuit to a classical processor, further reducing noise effects. Finally, CAFQA~\cite{ravi2022cafqa} accelerates convergence of variational quantum algorithms by searching over Clifford circuits to identify high-quality initial parameters, yielding substantial accuracy gains.

\section{Conclusion and Discussion}
\label{sec:discussion}

We have introduced ExtraFerm, a quantum circuit simulator tailored to particle number-conserving matchgates and controlled-phase gates. We have seen that ExtraFerm can offer significant advantages in latency and exponential advantages in memory consumption over tensor network and state vector methods. These advantages apply when specific Born-rule probabilities are required rather than the circuit's entire output distribution and when the circuit has low extent or sufficiently few controlled-phase gates.

Most notably, our integration of ExtraFerm with SQD improves the quality of molecular ground-state energy estimates with almost no computational overhead. We anticipate two avenues for future work. 1) Our simple heuristic of choosing high-probability bitstrings to mitigate error is effective, but more sophisticated uses of ExtraFerm that incorporate richer chemical insights may yield further improvements. 2) ExtraFerm could be used as a subroutine when optimizing the parameters of a quantum circuit with a gradient-free method such as simultaneous perturbation stochastic approximation (SPSA). More generally, we hope that researchers working with matchgate-based circuits will find ExtraFerm useful and discover new ways to use it to advance quantum computational chemistry.

\bibliographystyle{IEEEtran}
\bibliography{references}

\end{document}